\documentclass[english]{article}
\usepackage[T1]{fontenc}
\usepackage[latin9]{inputenc}
\usepackage{lmodern}
\usepackage{geometry}
\usepackage{babel}
\usepackage{float}
\usepackage{amsmath}
\usepackage{graphicx}
\usepackage{setspace}
\usepackage[unicode=true]
 {hyperref}
\usepackage{natbib}
\bibpunct{[} {]} {,} {n} {,} {,}											% formats bibliography punctuation



\begin{document}

\title{\textbf{Thermodynamic Volume of  Kerr--bolt--AdS Spacetime\vspace{0.75 in}}}

\author{Scott MacDonald\\
\\
\emph{Department of Physics and Astronomy}\\
\emph{University of Southern California}\\
\emph{Los Angeles, CA 90089-0484, U.S.A.}\\
\emph{}\\
smacdona {[}at{]} usc.edu}
\date{}
\maketitle
\vspace{0.5 in}
\begin{abstract}
In theories of gravity where the cosmological constant defines a thermodynamic
variable, the pressure, it has been shown that solutions of Einstein's
equations have a corresponding thermodynamic volume. In general, the
expression for the volume is not the same as the one arising from
naive geometrical considerations. Both rotation and nut charge are
properties known in the literature to give non-geometric thermodynamic
volumes, and so we combine the two in a single example and compute
the volume for the Kerr--bolt--AdS spacetime, presenting a new expression
that generalizes the previously known non-geometric cases.
\end{abstract}
\pagebreak{}

\section{Introduction}

Black hole thermodynamics \cite{Bekenstein:1973ur, Bekenstein:1974ax, Hawking:1974sw, Hawking:1976de}
is a subject long known for leading to deep insights into the nature
of both classical and quantum gravity. In the last few years the standard
story of how one associates gravitational quantities with thermodynamic
ones has been extended%
\footnote{For a selection of references where this has been explored, see refs.
\cite{Caldarelli:1999xj, Wang:2006eb, Sekiwa:2006qj, LarranagaRubio:2007ut, Kastor:2009wy, Dolan:2010ha, Cvetic:2010jb, Dolan:2011jm, Dolan:2011xt, Henneaux:1984ji, Teitelboim:1985dp, Henneaux:1989zc, Johnson:2014xza, Johnson:2014yja}.
See also the reviews in refs. \cite{Altamirano:2014tva, Dolan:2012jh}. For references where this has been explored in the context of conformal gravity and Lovelock gravity, see refs. \cite{Xu:2014kwa} and \cite{Xu:2014tja}, respectively.%
}, by allowing the cosmological constant $\Lambda$ to be a thermodynamic
variable, the pressure, $p=-\Lambda/8\pi$ (we use units such that $G=c=\hbar=k_{B}=1$). In this extended thermodynamics,
instead of being the internal energy $U$, the mass $M$, is identified
\cite{Kastor:2009wy} with the enthalpy: $M=H\equiv U+pV$. 

Given the association of the pressure with the cosmological constant
one may wonder what the conjugate thermodynamic volume corresponds
to. It was shown in refs. \cite{Kastor:2009wy, Dolan:2010ha, Cvetic:2010jb} that for
static black holes with horizon radius $r_{h}$ the thermodynamic
volume $V$ is the geometric volume\footnote{A definition for the volume of a stationary black hole was proposed in \cite{Parikh:2005qs}, and they obtain this same result for static holes, but for non-static black holes, the expression they find is different.} $V_{geo}=\frac{4}{3}\pi r_{h}^{3}$. However, this is not true in general; for example, for the Kerr--AdS
black holes the thermodynamic volume and the geometric volume are
not the same \cite{Cvetic:2010jb}. More recently, ref. \cite{Johnson:2014xza}
proposed that the extended thermodynamics should hold not just for
black hole spacetimes but for \emph{any} spacetime. That is, one should
associate the mass with the enthalpy in any system where the cosmological
constant is dynamical. As examples, the Taub--NUT--AdS and Taub--bolt--AdS
spacetimes were shown to also be cases where the thermodynamic volume
is different from the geometric one. The Taub--NUT--AdS spacetime
is an extreme example of this, because while the geometric volume
vanishes, the thermodynamic volume is finite.

In light of the work in refs. \cite{Cvetic:2010jb, Johnson:2014xza}, it is interesting
to explore more examples in which the thermodynamic volume and the
geometric one are not the same. Having seen two examples in the literature
where this is the case --- Taub--NUT/bolt--AdS and the Kerr--AdS black
hole --- we consider here a spacetime that is a mixture of the two.
We explore the impact of having both rotation (characterized by rotation
parameter $a$) and nut charge (characterized by $n$) in order to
try and better understand their contribution in the thermodyanmic
volume. To this end we study the Kerr--bolt--AdS spacetimes in four
dimensions%
\footnote{As we review in the next section, there are no Kerr--NUT--AdS spacetimes.%
} \cite{Mann:1999bt}.

The rest of the paper is organized as follows: we first present the
Kerr--bolt--AdS spacetime solutions, which have both rotation and
nut charge, along with some of their known thermodynamics. Then using
the extended thermodynamics, taking $p$ as dynamical and associating
the mass with the enthalpy, we compute the thermodynamic volume $V$.
The paper ends with a discussion of our results.

\section{The Geometry}

\noindent The metric for the (Euclideanized) Kerr--bolt--AdS spacetime
takes the form \cite{Mann:1999bt}:
\begin{align}
ds^{2} & =\frac{\mathcal{V}(r)(d\tau-(2n\cos\theta-a\sin^{2}\theta)d\phi)^{2}+\mathcal{H}(\theta)\sin^{2}\theta(ad\tau-(r^{2}-n^{2}-a^{2})d\phi)^{2}}{\chi^{4}(r^{2}-(n+a\cos\theta)^{2})}\\
 & +(r^{2}-(n+a\cos\theta)^{2})\left(\frac{dr^{2}}{\mathcal{V}(r)}+\frac{d\theta^{2}}{\mathcal{H}(\theta)}\right),\nonumber 
\end{align}
where the metric functions are given by
\begin{align}
\mathcal{H}(\theta) & =1+\frac{qn^{2}}{l^{2}}+\frac{(2n+a\cos\theta)^{2}}{l^{2}},\\
\mathcal{V}(r) & =\frac{r^{4}}{l^{2}}+\frac{((q-2)n^{2}-a^{2}+l^{2})r^{2}}{l^{2}}-2mr-\frac{(a^{2}-n^{2})(qn^{2}+l^{2}+n^{2})}{l^{2}}.
\end{align}

The parameter $n$ is the nut charge, $m$ the mass parameter, and
$a$ the rotation parameter. The cosmological constant, $\Lambda$,
sets the length scale $l$ via $\Lambda=-3/l^{2}$. The parameters
$q$ and $\chi$ as well as the periodicity of $\tau$ are fixed by
removing the conical singularities where the $\tau$ fibre degenerates.
Requiring smoothness of the metric in the $(\theta,\phi)$ section
leads to $q=-4$ and $\chi=\sqrt{1+a^{2}/l^{2}}$. Requiring smoothness
in the $(r,\tau)$ section implies that $\tau$ has period given by
$2\pi/\kappa$ with%
\footnote{Note that formula (4) differs from that of ref. \cite{Mann:1999bt} by
a factor of $2\pi$ in the denominator and that $\chi$ in ref. \cite{Mann:1999bt}
is 1 over what we have here (which matches refs. \cite{Ghosh:2003yx}
and \cite{Ghezelbash:2009gy}). %
}
\begin{equation}
\kappa=\frac{\mathcal{V}'(r_{+})}{2\chi^{2}\left(r_{+}^{2}-r_{n}^{2}\right)}.
\end{equation}

Here $r_{+}$ is the location where the foliation breaks down, i.e.,
$\mathcal{V}(r_{+})=0$ and $r_{n}=\sqrt{a^{2}+n^{2}}$ is the location
of the nut, where the area of the surfaces orthogonal to the $(r,\tau)$
section vanishes \cite{Mann:1999bt}. The metric reduces to both the
Kerr--AdS black hole and Taub--bolt--AdS when $n=0$ and $a=0$, respectively%
\footnote{See the discussion in the next section about matching conventions.%
}.

There is subtlety in fixing the periodicity of $\tau$ however, because
regularity along the Misner string \cite{Misner:1963fr} singularity, which
runs along the $z$-axis from the nut to infinity, requires $\tau$
to have period $8\pi n$. The resolution is to use $\mathcal{V}(r_{+})=0$
to determine the mass parameter $m$ in terms of $r_{+}$, and then
equate the two periods to give a quartic contraint on $r_{+}$ in
terms of $a,\, n$ and $l$. The result for $r_{+}$ is given in the
Appendix, and the expression for $m$ from the constraint is
\begin{equation}
m=\frac{r_{+}^{2}\left(-a^{2}+l^{2}-6n^{2}\right)+\left(n^{2}-a^{2}\right)\left(l^{2}-3n^{2}\right)+r_{+}^{4}}{2l^{2}r_{+}}.
\end{equation}
We will take the period of $\tau$ to be $2\pi/\kappa$ for computing
the temperature. One might be curious what happens when $r_{+}=r_{n}$,
the analogue of the NUT solution in the non-rotating case. Regularity
of the solution demands that $\mathcal{V}(r)$ have a double root,
but this requirement turns out to be incompatible with the periodicity
constraints unless $a=0$, thus $r_{+}>r_{n}$ (there are no regular
Kerr--NUT or Kerr--NUT--AdS solutions \cite{Mann:1999bt, Gibbons:1979nf})%
\footnote{Sometimes in the literature the metric (1) is reffered to as Kerr--NUT--AdS,
but here we differentiate NUT vs. bolt and so refer to the spacetime
as Kerr--bolt--AdS.%
}.

\section{Gravitational Thermodynamics and the Volume}

We are now in a position to do the gravitational thermodynamics. The
on-shell (Euclidean) action was found in ref. \cite{Mann:1999bt}, and
the result is
\begin{equation}
I=-\pi\frac{\left(r_{+}^{2}-n^{2}-a^{2}\right)\left[r_{+}^{4}-(a^{2}+l^{2})r_{+}^{2}+(n^{2}-a^{2})(3n^{2}-l^{2})\right]}{\left(3r_{+}^{4}+(l^{2}-a^{2}-6n^{2})r_{+}^{2}+(n^{2}-a^{2})(3n^{2}-l^{2})\right)\chi^{2}}.
\end{equation}

Using the thermodynamic relations we can compute the entropy from
this expression via $S=\beta\partial_{\beta}I-I=\frac{1}{T}\left(\overline{M}+\overline{\Omega}J\right)-I$
where the inverse temperature $\beta=1/T=2\pi/\kappa$, $\overline{M}$
is the energy of the spacetime at infinity (the mass), $J$ is the
conserved charge associated with the Killing vector $\partial/\partial\phi$,
and $\overline{\Omega}$ is the angular velocity of the horizon measured
in a frame rotating at infinity, given by $\overline{\Omega}=a/\left(r_{+}^{2}-a^{2}-n^{2}\right)$.
Both $\overline{M}$ and $J$ were found in ref. \cite{Mann:1999bt}
to be $\overline{M}=\frac{m}{\chi^{4}}$ and $J=\frac{ma}{\chi^{4}}$,
and so we can use these plus the expression for $m$ in (5) to compute
the entropy, which is
\begin{equation}
S=\frac{\pi\left(r_{+}^{4}\left(-4a^{2}+l^{2}-15n^{2}\right)+\left(n^{2}-a^{2}\right)^{2}\left(3n^{2}-l^{2}\right)+r_{+}^{2}\left(a^{2}+3n^{2}\right)^{2}+3r_{+}^{6}\right)}{\chi^{2}\left(r_{+}^{2}\left(-a^{2}+l^{2}-6n^{2}\right)+\left(n^{2}-a^{2}\right)\left(3n^{2}-l^{2}\right)+3r_{+}^{4}\right)}.
\end{equation}

There is a subtlety here pointed out in ref. \cite{Gibbons:2004ai}. If
one computes the angular velocity $\overline{\Omega}$ in a frame
that is rotating at infinity then the First Law of Thermodynamics
does not hold. Instead, one needs $\overline{\Omega}$ computed in
a frame that is non--rotating at infinity, which is found by taking
$\phi\rightarrow\phi+l^{-2}a\tau$ in (1). This shift changes both
$\overline{\Omega}$ and $\overline{M}$, and gives the relationship
between the rotating and non--rotating frame quantities to be 
\begin{align}
\overline{\Omega} & =\Omega+\frac{a}{l^{2}},\\
\overline{M} & =M+\frac{a}{l^{2}}J.
\end{align}

Now following refs. \cite{Kastor:2009wy, Cvetic:2010jb, Johnson:2014xza} we take the
cosmological constant of the spacetime to supply a thermodynamic pressure,
$p=-\Lambda/8\pi=3/\left(8\pi l^{2}\right)$, and label its conjugate
volume $V$. In this extended thermodynamics the mass $M$, instead
of being the internal energy $U$, is identified with the enthalpy,
$M=H\equiv U+pV$. The First Law is then 
\begin{equation}
dH=TdS+\Omega dJ+Vdp.
\end{equation}

One could proceed to find the volume using the thermodynamic relation
$V=\left(\partial H/\partial p\right)\left|_{S,J}\right.$. Instead,
we simplify matters by using a Smarr relation \cite{Caldarelli:1999xj, Kastor:2009wy, Cvetic:2010jb, Johnson:2014xza, Gibbons:2004ai, Smarr:1972kt}
which follows entirely from scaling arguements:
\begin{equation}
\frac{H}{2}-TS-\Omega J+pV=0.
\end{equation}
We note that to find the transformation of $V$ between the non--rotating
and rotating frame we can use (8) and (9) in (11) to find $\overline{V}=V+\frac{a}{2l^{2}p}J$. 

From (11) we find

\begin{alignat}{1}
V & =\frac{2\pi l^{2}\left(r_{+}^{6}\left(a^{2}-2l^{2}\right)+r_{+}^{4}\left(-2a^{4}+a^{2}\left(l^{2}-7n^{2}\right)+8l^{2}n^{2}\right)\right)}{3r_{+}\left(a^{2}+l^{2}\right)^{2}\left(a^{2}+n^{2}-r_{+}^{2}\right)}\\
 & +\frac{2\pi l^{2}\left(r_{+}^{2}\left(a^{6}+10a^{4}n^{2}+a^{2}\left(-4l^{4}+16l^{2}n^{2}+3n^{4}\right)-6l^{2}n^{4}\right)+a^{2}(a^{2}-n^{2})\left(l^{2}-3n^{2}\right)\left(a^{2}+4l^{2}+n^{2}\right)\right)}{3r_{+}\left(a^{2}+l^{2}\right)^{2}\left(a^{2}+n^{2}-r_{+}^{2}\right)},\nonumber 
\end{alignat}
where $r_{+}$ is given in the Appendix.

Before proceeding to understand this result, let us check some special
cases. Putting $a=0$ above we find that 
\begin{equation}
V|_{a=0}=\frac{4}{3}\pi\left(r_{b}^{3}-3n^{2}r_{b}\right),
\end{equation}
 where $r_{b}=r_{+}|_{a=0}$ -- see the Appendix. This is the volume
of Taub--bolt--AdS found in ref. \cite{Johnson:2014xza}. We note that in
the metric (1), we recover the metric used in ref. \cite{Johnson:2014xza}
after changing the sign of the azimuthal coordinate $\phi\rightarrow-\phi$.

Putting $n=0$ we recover the Kerr--AdS black hole metric, thermodynamics\footnote{We point out that the gravitational and thermodynamic stability of Kerr--AdS black holes in five and higher dimensions was studied in \cite{Carter:2005uw} but with no pressure term. }
and volume as in ref. \cite{Cvetic:2010jb}: 
\begin{equation}
V|_{n=0}=\frac{2\pi l^{2}\left(-r_{+}^{4}\left(a^{2}-2l^{2}\right)+a^{2}r_{+}^{2}\left(a^{2}+l^{2}\right)+a^{2}l^{2}\left(a^{2}+4l^{2}\right)\right)}{3\left(a^{2}+l^{2}\right)^{2}r_{+}},
\end{equation}
were $r_{+}$ is now dependent on the mass parameter $m$ (see Appendix),
as there is no additional regularity constraint to eliminate $m$
as an independent parameter. Care must be taken in comparing (14)
with the results in ref. \cite{Cvetic:2010jb} due to different conventions.
In particular, the $\tau$ coordinate in (1) is given by $(1-\tilde{a}{}^{2}\tilde{l}^{-2})\tilde{\tau}$
where a tilde denotes their symbols, and also one must take $\tilde{a}\rightarrow ia$,
which comes from Euclideanisation (and so one must be careful in this
when comparing the angular velocities which are to be computed in
Lorentzian signature), and $\tilde{\Lambda}\rightarrow-\Lambda$ which
comes from a difference in the sign of the cosmological constant term
in the action%
\footnote{In ref. \cite{Cvetic:2010jb} they choose to work with thermodynamic variables
$\Lambda$ and $\Theta$, instead of $p$ and $V$. When matching
their formulae, one must also take $V\rightarrow+8\pi\Theta$%
}.

\section{Exploring the Volume Formula}

\begin{figure}[t]
\begin{centering}
\includegraphics[scale=0.53]{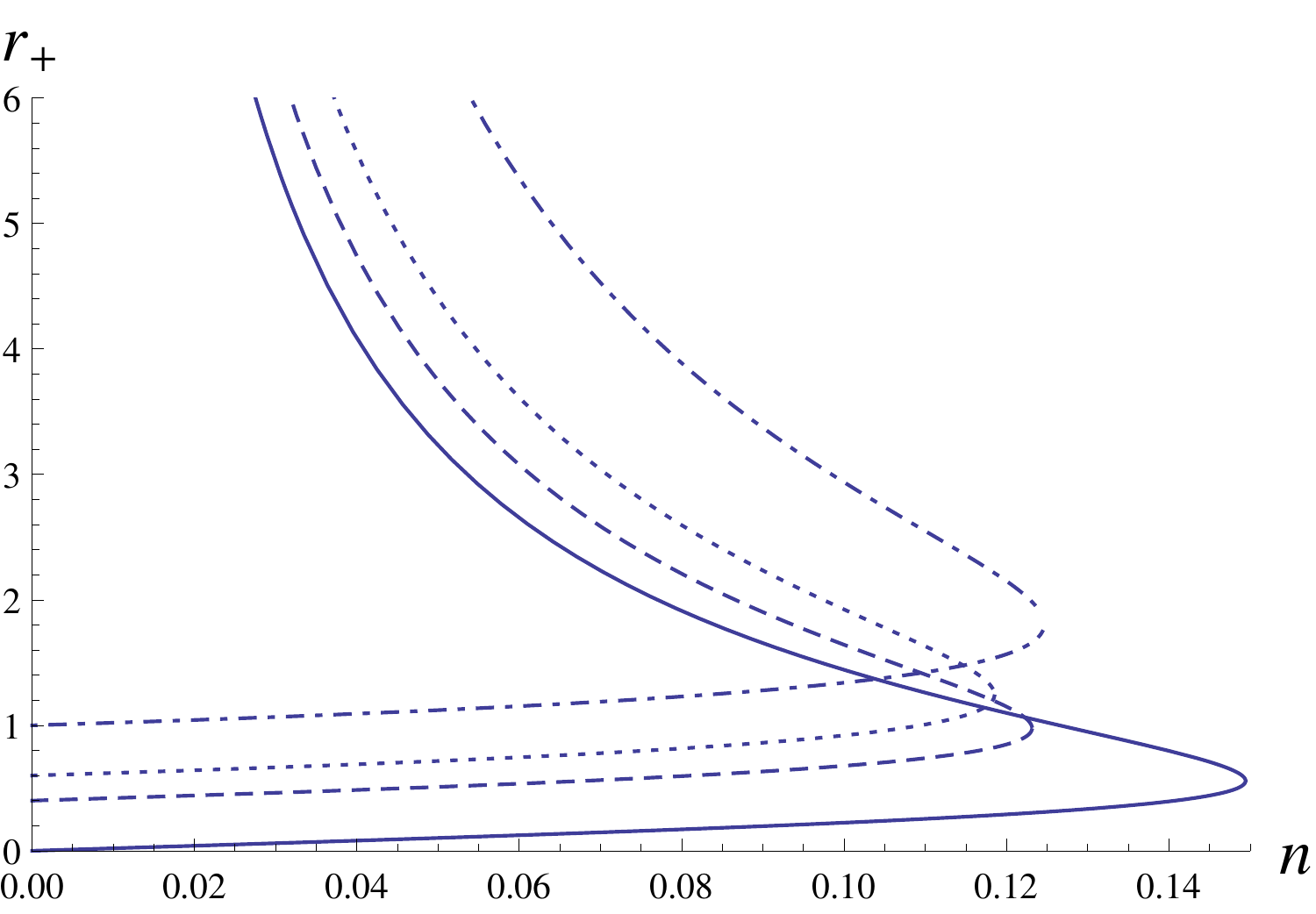}\includegraphics[scale=0.53]{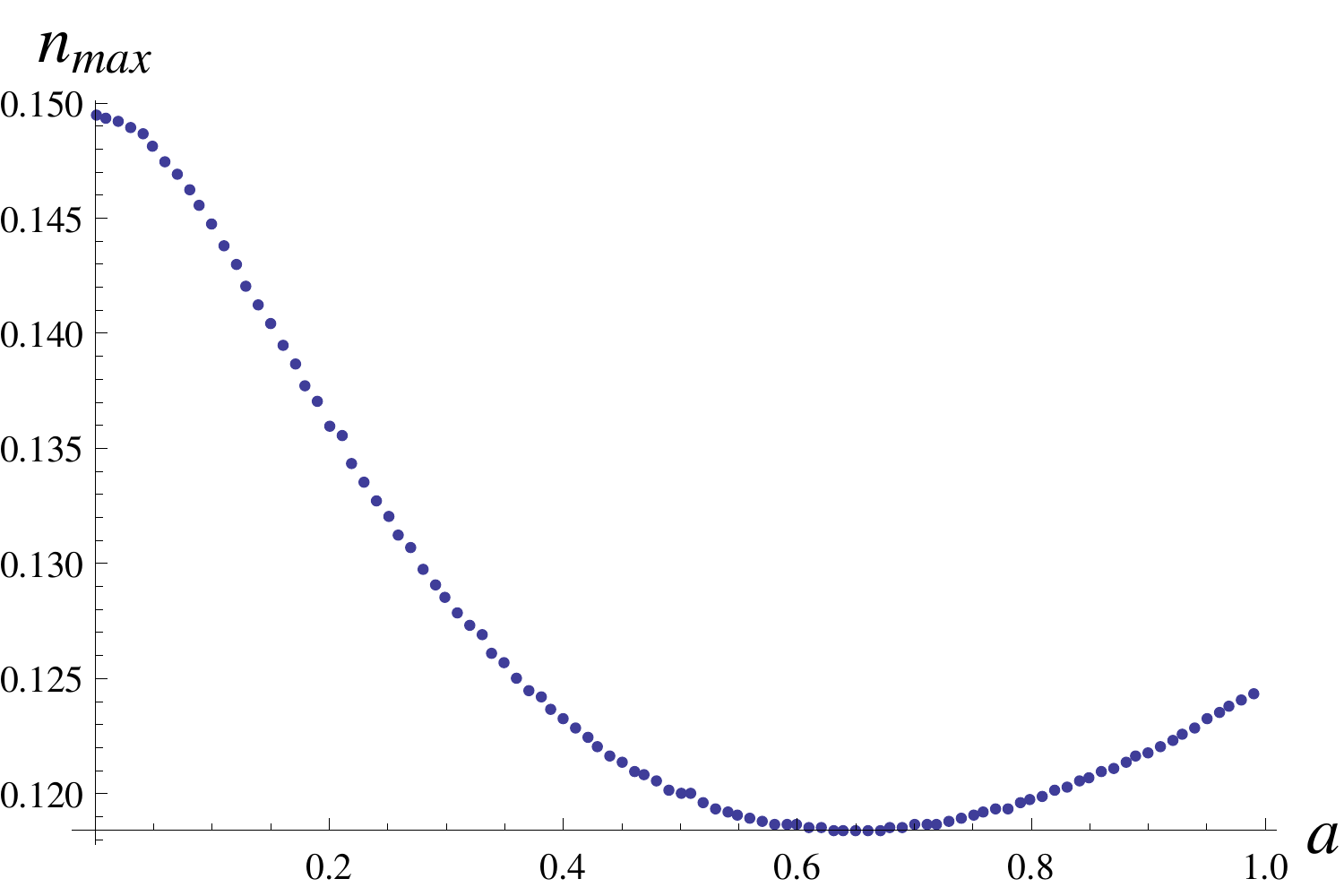}
\par\end{centering}

\caption{$r_{+}$ as a function of $n$. The (approximate) maximum value of
$n_{max}$ depends on $a$ (taking $l=1$), and we connect the upper
and lower branch for $r_{+}$ at this point. The curves correspond
to: solid $a=0$, dashed $a=0.4$, dotted $a=0.6$, dot-dashed $a=1.0$.
Also shown is the value of $n_{max}$ for various values of $a$. }
\end{figure}
With (12) in hand we can begin to explore its behavior with respect
to the various parameters we have: $a,\, n,\, r_{+}$ and $l$. First,
however, we need to understand the allowed ranges of the parameters.
We plot in Figure 1 $r_{+}$ as a function of $n$ for various values
of $a$, setting $l=1$. It is the ratio $a/l$ that is relevant,
but for simplicity we have taken $l=1$ and so allow $0<a<1$. 

Note that if one writes the metric (1) in Lorentzian signature (or
rather, derives (1) as the Euclidean signature version) then one must
take $a\rightarrow i\alpha$ and so, in particular, $\chi\rightarrow\sqrt{1-\alpha^{2}/l^{2}}$
and we see that when $\alpha^{2}=l^{2}$ the metric becomes singular
\cite{Hawking:1998kw}, thus the rotation parameter is restricted to $\alpha^{2}<l^{2}$.
Additionally, the allowed values for $r_{+}$ are such that it remains
real and greater than $r_{n}=\sqrt{a^{2}+n^{2}}$, meaning that there
is a certain maximum value $n$, denoted $n_{max}$. In Figure 1 we
have plotted the approximate value of $n_{max}$ for various values
of $a$, found by numerically scanning for the value of $n$ for which
$\text{Im}(r_{+}(n))$ first becomes non--zero. To get the full behavior
of $r_{+}(n)$ we must recall that $r_{+}$ in general satifies a
quartic constraint and we take the two largests roots, where the cross--over
between the two branches occurs at $n_{max}$. We note how $n_{max}$
decreases as a function of $a$ for $a<0.65$, but increases as a
function of $a$ for $a>0.65$. 

Using this we turn to plotting the volume as a function of $n$ for
various values of $a$, and taking $l=1$ --- the result is in Figure
2 (a). We see that the volume appears to be a smooth function of $n$
and positive--definite. The turning point of $V$ corresponds to when
$n$ has reached $n_{max}$ for that particular value of $a$. We
can also look at the behavior of $V$ as a function of $l$. In the
case when $n=0$ we have plotted the result in Figure 2 (b). We find
again that the volume appears to be positive--definite. It is difficult
to show analytically that the volume is always positive--definite,
but our plots strongly indicate this to be true%
\footnote{Positive--definiteness is not guaranteed for the thermodynamic volume
of a spacetime. For an example of a negative volume, see ref. \cite{Johnson:2014xza}
and the interpretation therein. Also, in ref. \cite{Xu:2013zea}, a negative thermodynamic volume for a black hole in Gauss-Bonnet gravity was found.%
}.

\begin{figure}[H]
\begin{centering}
\includegraphics[scale=0.9]{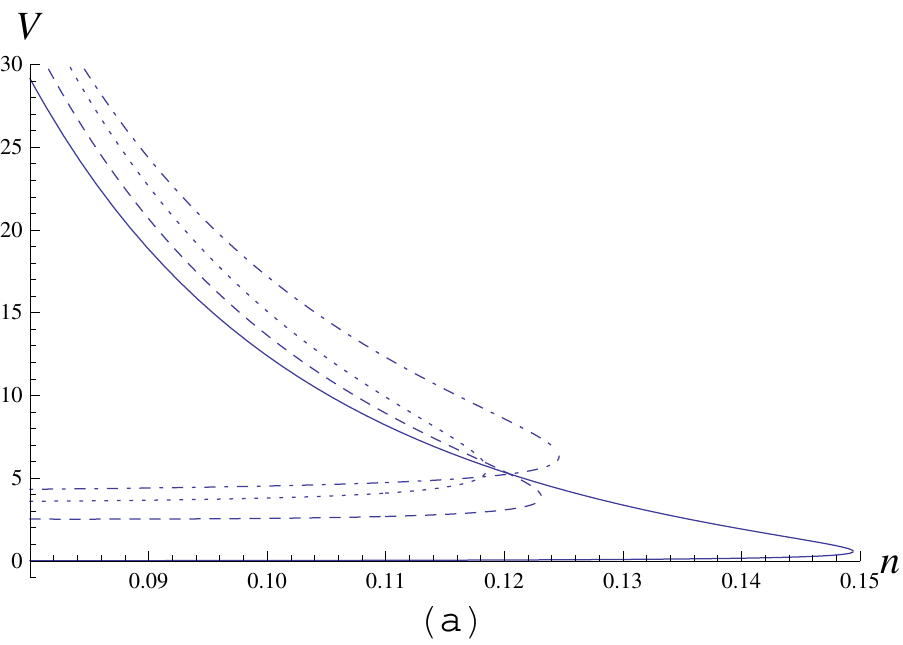}\includegraphics[scale=0.9]{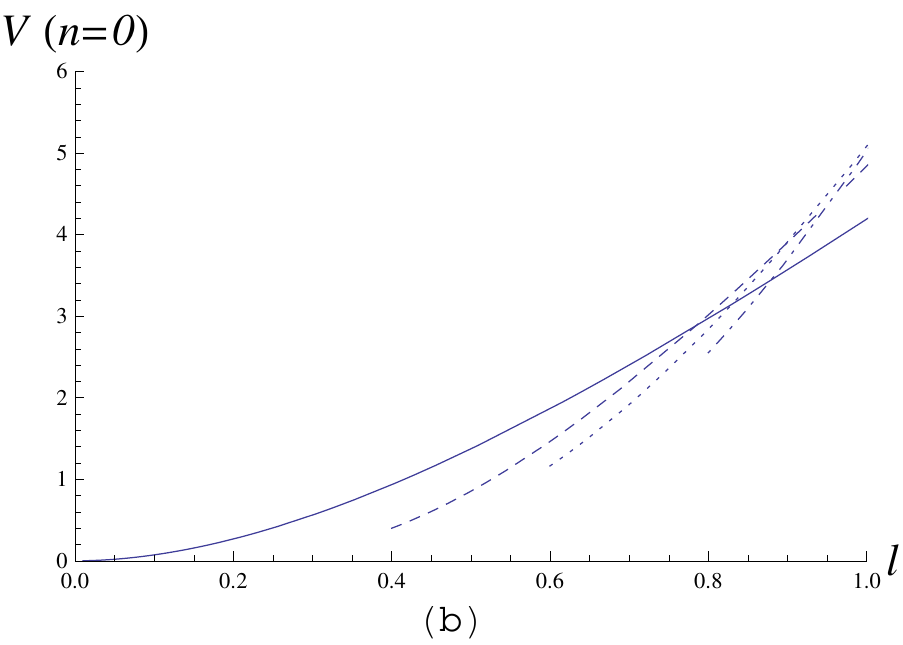}
\par\end{centering}

\caption{(a) $V$ as a function of $n$. The curves correspond to: solid $a=0$,
dashed $a=0.4$, dotted $a=0.6$, dot-dashed $a=1.0$. (b) $V$ as
a function of $l$ when $n=0$. Here $r_{+}$ depends on $m$ which
is an independent variable. We have taken $m=1$. The curves correspond
to: solid $a=0$, dashed $a=0.4$, dotted $a=0.6$, dot-dashed $a=0.8$.}
\end{figure}

One observation is that the thermodynamic volume is not just the sum
of the volumes for the Kerr--AdS and Taub--bolt spacetime
\begin{equation}
V\neq V|_{a=0}+V|_{n=0}.
\end{equation}
We can also expand the volume about both small $a$ and small $n$.
We consider the former first. We make a series expansion for $m$
about $a=0$ and use the condition $\mathcal{V}(r_{+})=0$ to determine
the coefficients; to second order in $a$ we find 
\begin{align}
m & =\sum_{i=0}^{\infty}f_{i}a^{i}=f_{0}+f_{1}a+f_{2}a^{2}+\mathcal{O}(a^{3}),\\
f_{0} & =\frac{l^{2}n^{2}-3n^{4}+l^{2}r_{+}^{2}-6n^{2}r_{+}^{2}+r_{+}^{4}}{2l^{2}r_{+}},\quad\quad f_{1}=0,\nonumber \\
f_{2} & =\frac{3n^{2}-l^{2}-r_{+}^{2}}{2l^{2}r_{+}}.\nonumber 
\end{align}
Using the condition that $8\pi n=2\pi/\kappa$ (with $\kappa$ given
in (4)) we can find a series expansion for $r_{+}$ about $a=0$.
It becomes rather messy at second order so we show it here only to
first order:
\begin{align}
r_{+} & =\sum_{i=0}^{\infty}g_{i}a^{i}=g_{0}+g_{1}a+\mathcal{O}(a^{2}),\\
g_{0} & =r_{b}\equiv\frac{l^{2}}{12n}\left(1\pm\sqrt{1-48\frac{n^{2}}{l^{2}}+144\frac{n^{4}}{l^{4}}}\right),\quad\quad g_{1}=0.\nonumber 
\end{align}
where $r_{b}=r_{+}|_{n=0}$ is just the solution to the quadratic
equation for the Taub--bolt--AdS spacetime \cite{Johnson:2014xza}. With
(16) and (17) we can expand the volume to first order in $a$ and
find
\begin{align}
V & \approx\frac{4}{3}\pi\left(g_{0}^{3}-3n^{2}g_{0}\right)+4\pi\left(g_{0}^{2}-n^{2}\right)g_{1}a+\mathcal{O}(a^{2})\\
 & =\frac{4}{3}\pi\left(r_{b}^{3}-3n^{2}r_{b}\right)+\mathcal{O}(a^{2}).\nonumber 
\end{align}
which tells us that the volume is just that of Taub--bolt--AdS \cite{Johnson:2014xza}
to first order in $a$.

We may repeat the above proceedure for small $n$, which turns out
to be more interesting. We find from $\mathcal{V}(r_{+})=0$ that
\begin{align}
m & =\sum_{i=0}^{\infty}h_{i}n^{i}=h_{0}+h_{1}n+h_{2}n^{2}+\mathcal{O}(n^{3}),\\
h_{0} & =\frac{-a^{2}l^{2}-a^{2}r_{+}^{2}+l^{2}r_{+}^{2}+r_{+}^{4}}{2l^{2}r_{+}},\quad\quad h_{1}=0,\nonumber \\
h_{2} & =\frac{3a^{2}+l^{2}-6r_{+}^{2}}{2l^{2}r_{+}}.\nonumber 
\end{align}
Using the condition $8\pi n=2\pi/\kappa$ and writing 
\begin{equation}
r_{+}=\sum_{i=0}^{\infty}k_{i}n^{i},
\end{equation}
we find that either $k_{0}=\pm a$ or $k_{0}=0$. The latter is inconsistent,
so we have (taking the positive sign)
\begin{align}
k_{0} & =a,\quad\quad k_{1}=2,\quad\quad k_{2}=\frac{29a^{2}-3l^{2}}{2a(a^{2}+l^{2})}.
\end{align}
From (19) and (20) we find $V$ to first order in small $n$ to be
\begin{equation}
V\approx\frac{8\pi al^{4}}{3\left(a^{2}+l^{2}\right)}-\frac{4\pi\left(2a^{4}l^{2}-5a^{2}l^{4}+3l^{6}\right)}{3\left(a^{2}+l^{2}\right)^{2}}n+\mathcal{O}(n^{2}).
\end{equation}

We note that (22) at $0$th order is not (14), the result for the
Kerr--AdS black hole \cite{Cvetic:2010jb}. This is because small $n$
and $n=0$ are not the same. For small $n$ we still have to impose
the regularity condition due to the Misner string, and so we have
the $8\pi n=2\pi/\kappa$ constraint which allows us write $r_{+}$
in terms of $a,\, l$ and small $n$, obtaining (22) for the volume.
When $n=0$ there is no Misner string condition and so no additional
constraint on $r_{+}$. We write $V$ in terms of $m$ when $n=0$
and then substitute $m$ in terms of $r_{+},\, a$ and $l$ (which
is the $0$th order term in (19)) to obtain (14). We also note that
there is a first order contribution to the volume for small $n$ in
contrast to the small $a$ expansion (18), in which $a$ doesn't appear
at first order.

\section{Conclusion}

We have computed the thermodynamic volume for a class of Kerr--bolt--AdS
spacetimes and explored its behavior as a function of the rotation
parameter $a$ and the nut charge $n$. Our plots in Figure 2 strongly
suggest that this volume is positive--definite, as expected. Additionally
we found that the thermodynamic volumes for the two limiting cases
($a=0$ and $n=0$) do not simply add to give $V$. Taking the small
$a$ expansion we found that $a$ enters the volume at second order,
in contrast to the small $n$ expansion in which there is a first
order contribution from $n$. 

There are many future directions of research one can explore. One
possibility would be to extend our analysis to higher dimensions,
though care has to be taken when dealing with even vs. odd dimension.
It is interesting to continue to study other examples where the thermodynamic
and the geometric volume differ, and perhaps eventually classify the
various characteristic that go into making the thermodynamic volume.
There is also the matter of understanding and interpreting the thermodynamic
volume in the context of gauge/gravity duality, so one could consider
our thermodynamics in the Kerr/CFT correspondence \cite{Ghezelbash:2009gy, Guica:2008mu}
to see what could be learned. We leave these matters for future investigation.

\section*{Acknowledgements}

Scott MacDonald would like to thank Clifford Johnson for his guidance
and wisdom during this project, and his comments on the manuscript.
He would also like to thank the US Department of Energy for support
under grant DE-FG03-84ER-40168, the USC Dornsife College of Letters,
Arts, and Sciences, and Jessica for her support and encouragement.

\section*{Appendix}

For completeness we give the full expression for $r_{+}$, which is
constrained by the regularity condition explained in Section 2 to
be a solution of the quartic polynomial defined by $\beta=1/T=8\pi n=\kappa/2\pi$,
with $\kappa$ given in (4). We can write this constraint in the standard
form
\[
Ar_{+}^{4}+Br_{+}^{3}+Cr_{+}^{2}+Dr_{+}+E=0,
\]
for which known formulae exist for the solution, and the roots may
be written
\begin{align*}
r_{+\,1,2} & =-\frac{B}{4A}+Z\pm\frac{1}{2}\sqrt{-4Z^{2}-2W-\frac{Y}{Z}},\\
r_{+\,3,4} & =-\frac{B}{4A}-Z\pm\frac{1}{2}\sqrt{-4Z^{2}-2W+\frac{Y}{Z}},
\end{align*}
where we take $r_{+\,1,2}$ as the two branches of $r_{+}$ in the
paper and
\begin{align*}
W & =\frac{8AC-3B^{2}}{8A^{2}},\quad\quad Y=\frac{B^{3}-4ABC+8A^{2}D}{8A^{3}},\\
Z & =\frac{1}{2}\sqrt{-\frac{2}{3}W+\frac{1}{3A}\left(Q+\frac{\Delta_{0}}{Q}\right)},\quad\quad Q=\sqrt[3]{\frac{\Delta_{1}+\sqrt{\Delta_{1}^{2}-4\Delta_{0}^{3}}}{2},}\\
\Delta_{0} & =C^{2}-3BD+12AE,\\
\Delta_{1} & =2C^{3}-9BCD+27B^{2}E+27AD^{2}-72ACE.
\end{align*}
For Kerr--bolt--AdS spacetime the coefficients of the quartic for
$r_{+}$ are
\begin{align*}
A & =6n,\quad\quad\quad\quad\quad\quad\quad\quad B=-(a^{2}+l^{2}),\\
C & =2(l^{2}-a^{2}-6n^{2}),\quad\quad D=(a^{2}+l^{2})(a^{2}+n^{2}),\\
E & =2n(l^{2}-3n^{2})(a^{2}-n^{2}).
\end{align*}
Taking $a=0$ in the metric reduces the constraint on $r_{+}$ to
be quadratic and we find \cite{Johnson:2014xza} 
\[
r_{b}=r_{+}|_{a=0}=\frac{l^{2}}{12n}\left(1\pm\sqrt{1-48\frac{n^{2}}{l^{2}}+144\frac{n^{4}}{l^{4}}}\right).
\]
Taking $n=0$ in the metric we can no longer impose the condition
on $m$, so it must now be taken as an independent parameter and we
may write the quartic constraint on $r_{+}|_{n=0}$ in the standard
form above with
\begin{align*}
A & =\frac{1}{l^{4}},\quad\quad B=0,\\
C & =1-\frac{a^{2}}{l^{2}},\quad\quad D=-2m,\\
E & =-a^{2}.
\end{align*}

%\begin{thebibliography}{10}										Commented

%\bibliography{macdonaldbib}											% research is the .bib file name
%\bibliographystyle{utphys}										% style is h-physrev

\providecommand{\href}[2]{#2}\begingroup\raggedright\endgroup

\end{document}